\documentclass{osa-article}

\journal{oe}

\articletype{Research Article}
\usepackage[latin1]{inputenc}
\usepackage{lineno}
\linenumbers

\begin{document}

\title{Explanation of the anomalous redshift on nonlinear X-ray Compton scattering spectrum by a bound electron}
\author{Shang Shi,\authormark{1,7} Jing Chen,\authormark{2,3} Yujun Yang,\authormark{4} Zhong-Chao Yan,\authormark{5,6} Xiaojun Liu,\authormark{6} Bingbing Wang,\authormark{1,7,*}}

\address{\authormark{1}Laboratory of Optical Physics, Beijing National Laboratory for Condensed Matter Physics, Institute of Physics, Chinese Academy of Sciences, Beijing 100190, China\\
\authormark{2}HEDPS, Center for Applied Physics and Technology, Peking University, Beijing 100871, China\\
\authormark{3}Institute of Applied Physics and Computational Mathematics, Beijing 100088, China\\
\authormark{4}Insititute of Atomic and Molecular Physics, Jilin University, Changchun 130012, China\\
\authormark{5}Department of Physics, University of New Brunswick, Fredericton, New Brunswick, E3B 5A3, Canada\\
\authormark{6}State Key Laboratory of Magnetic Resonance and Atomic and Molecular Physics, Wuhan Institute of Physics and Mathematics,
Innovation Academy for Precision Measurement Science and Technology, Chinese Academy of Sciences, Wuhan 430071, China\\
\authormark{7}University of Chinese Academy of Sciences, Beijing 100049, China}

\email{\authormark{*}wbb@aphy.iphy.ac.cn} 



\begin{abstract}
Nonlinear Compton scattering is an inelastic scattering process where a photon is emitted due to the interaction between an electron and an
intense laser field. With the development of X-ray free-electron lasers, the intensity of X-ray laser is greatly enhanced, and the signal from X-ray nonlinear Compton scattering is no longer weak. Although the nonlinear Compton scattering by an initially free electron has been thoroughly investigated, the mechanism of nonrelativistic nonlinear Compton scattering of X-ray photons by bound electrons is unclear yet. Here, we present a frequency-domain formulation based on the nonperturbative quantum electrodynamic to study nonlinear Compton scattering of two photons off a bound electron inside an atom in a strong X-ray laser field. In contrast to previous theoretical works, our results clearly reveal the existence of anomalous redshift phenomenon observed experimentally by Fuchs \emph{et al.} (Nat. Phys. \textbf{11}, 964 (2015)) and suggest its origin as the binding energy of the electron as well as the momentum transfer from incident photons to the electron during the scattering process. Our work builds a bridge between intense-laser atomic physics and Compton scattering process that can be used to study atomic structure and dynamics at high laser intensities.
\end{abstract}

\section{Introduction}
The Compton effect is well known for proving the quantum hypothesis of light in 1923 experimentally \cite{compton1923quantum}. Afterwards the impulse approximation (IA) approach to Compton scattering on bound electrons, which considers the initial electron to be free with the momentum distribution of the bound state, was put forward by DuMond in his work on the scattering of photons from solids \cite{du1929compton,dumond1933linear}. Since the Compton profile is directly related to the electron momentum distribution, the electronic structure of atoms, molecules and solids can be probed by X- and $\gamma$-ray linear Compton scattering (LCS) \cite{cooper1971compton,ching1974band,laurent1978energy,jaiswal2007kinetically} and analyzed by IA. For LCS process, Eisenberger and Platzman verified the validity of nonrelativistic IA for doubly differential cross sections \cite{eisenberger1970compton}, and later Ribberfors extended IA to relativistic region \cite{ribberfors1975relationship}. Modification of IA and reexamination of the validity of IA have been the main interests of recent studies \cite{florescu2009k,lajohn2010low,pratt2010compton}.

With the first use of the Linac Coherent Light Source at the SLAC National Accelerator Laboratory in 2010~\cite {emma2010first}, the era of exploring the nonlinear interaction of ultrafast and ultra-intense X-rays with matters has begun. By using X-ray free-electron lasers~\cite{mitzner2008spatio, mcneil2010x, ishikawa2012compact, galayda2010x, ribic2012status, bostedt2013ultra}, people have observed for the first time extensive nonlinear phenomena at X-ray wavelengths, including the X-ray second harmonic generation
in diamonds~\cite{shwartz2014x}, two-photon absorption in the hard X-ray region~\cite{tamasaku2014x,ghimire2016nonsequential}, electron femtosecond response to an ultra-intense X-ray radiation~\cite{young2010femtosecond}, and nonlinear Compton scattering (NCS) of X-ray photons~\cite{fuchs2015anomalous}. Among them, the NCS is a particularly interesting phenomenon because the observed anomalous redshift of the scattered photon can be regarded as a breakdown of the widely-used IA theory for bound electrons.

As far as we know, there exist a few theoretical studies ~\cite{hopersky2015compton, krebs2019time, venkatesh2020simulation} devoted to the NCS processes involving bound electrons in recent years, but no theoretical work demonstrated the experimental observations by Fuchs \emph{et al.} \cite{fuchs2015anomalous}. For example, Krebs \emph{et al.}~\cite{krebs2019time} developed a nonperturbative approach based on the time-dependent Schr\"{o}dinger equation to investigate linear and nonlinear Compton scatterings of X-ray photons by atoms. However, their results were consistent with the predictions of the free-electron model and do not support the existence of the redshift found in \cite{fuchs2015anomalous}. More recently, Venkatesh and Robicheaux~\cite{venkatesh2020simulation}  claimed that their theoretical results exhibit a blueshift compared with the scattered photon energy predicted by the free-electron model during an NCS process. Therefore, the origin of the anomalous redshift phenomenon observed by Fuchs \emph{et al.} \cite{fuchs2015anomalous} is still an open question.

Motivated by the theoretical gap of the NCS mechanism, in this work we will apply the frequency-domain theory based on the nonperturbative quantum electrodynamic (QED) to study the NCS process of bound electrons. This theory has previously been successfully applied to recollision processes in strong laser fields~\cite{guo1989scattering,gao2000nonperturbative,wang2007frequency,
wang2012frequency}. The advantages of the QED method in treating the NCS could be shortly provided. Specifically, we will focus on the double differential probability (DDP) for the NCS process of a bound electron in an X-ray laser field. Our calculation will clearly demonstrate that in the DDP spectrum of the two-photon NCS, as the energy of the scattered photon increases, a redshift peak will appear, which is in contrast to the results by free-electron model~\cite{mackenroth2011nonlinear,seipt2011nonlinear,angioi2016nonlinear}, and other theoretical predictions~\cite{krebs2019time, venkatesh2020simulation}. Our theoretical results can be considered as the first qualitative confirmation of the measurement of Fuchs \emph{et al.}~\cite{fuchs2015anomalous}.

\section{Frequency-domain theory of NCS by bound electrons}

The frequency-domain theory is based on the nonperturbative quantum electrodynamic, where the laser-matter system can be regarded as an isolated one, hence the total energy of the system is conserved during the laser-matter interaction process and the formal scattering theory~\cite{gell1953formal} can be applied. In this theory, the incident laser field, as a part of the whole system, is regarded as a quantized field, and all dynamic processes are treated as quantum transitions between two states of the laser-matter system. In the following, we develop this theory to investigate the NCS by a bound electron in intense laser fields. Natural units $(\hbar=c=1)$ are used throughout unless otherwise stated. The $T$-matrix element between the initial state $|\psi_i\rangle$ and the final state $|\psi_f\rangle$ is
\begin{equation}\label{eq9}
T_{fi}=\langle\psi_f|V|\psi_i\rangle+\langle\psi_f|V\frac{1}{E_i-H_0-U-V+i\varepsilon}V|\psi_i\rangle,
\end{equation}
where $H_0$ is the non-interaction part of the Hamiltonian for the atom-radiation system, $U$ is the atomic binding potential, and $V$ is the interaction operator between electron and photons. The initial state is $|\psi_i\rangle=\Phi_i(\textbf{r})\otimes|l\rangle\otimes|0\rangle$ with energy $E_i=(-E_B)+(l+\frac{1}{2})\omega_1+\frac{1}{2}\omega_2$ for $\omega_1$ the incident laser frequency and $\omega_2$ the scattered photon frequency. Here, $\Phi_i(\textbf{r})$ is the ground-state wave function of the atomic electron with the binding energy $E_B>0$, and $|l\rangle$ and $|0\rangle$ are the Fock states of the incident and scattered photons with photon number $l$ and 0, respectively. The final state is $|\psi_f\rangle=\Psi_{\textbf{P}_fn_f}\otimes|1\rangle$ with total energy $E_f=\textbf{P}_f^2/(2m)+(n_f+\frac{1}{2}+u_p)\omega_1+\frac{3}{2}\omega_2$, where $\Psi_{\textbf{P}_fn_f}$ is the Volkov state of the electron in the incident laser field~\cite{guo1989scattering} with $\textbf{P}_f$ being the final momentum of the electron and $u_p$ being the ponderomotive energy in unit of the incident photon energy.

The first and second terms in Eq.~(\ref{eq9}) correspond to a one-step and two-step transition, respectively. In this work, since the contribution of the two-step transition is much smaller than that of the one-step transition under the present laser conditions at the scattered photon energy around twice of the incident photon energy, the second term in Eq.~(\ref{eq9}) is dropped here and will be investigated in the future. Hence, the $T$-matrix element for NSC can be expressed as
\begin{equation}\label{eq10}
T_{fi}=T_{AP}+T_{AA-}+T_{AA+}\,,
\end{equation}
where $T_{AP}=\langle\psi_f|V_{22}|\psi_i\rangle$, $T_{AA-}=\langle\psi_f|V_{21-}|\psi_i\rangle$, and $T_{AA+}=\langle\psi_f|V_{21+}|\psi_i\rangle$. Here, the electron-photon interaction operators include $V_{22}=\frac{e}{m}g_2e^{-i\textbf{k}_2\cdot \textbf{r}}a_2^\dag\boldsymbol{\epsilon}_2^*\cdot (-i\nabla)$, $V_{21-}=\frac{e^2}{m}g_1g_2\boldsymbol{\epsilon}_2^*\cdot \boldsymbol{\epsilon}_1e^{i(\textbf{k}_1-\textbf{k}_2)\cdot \textbf{r}}a_1a_2^\dag$, and $V_{21+}=\frac{e^2}{m}g_1g_2\boldsymbol{\epsilon}_2^*\cdot \boldsymbol{\epsilon}_1^*e^{-i(\textbf{k}_1+\textbf{k}_2)\cdot \textbf{r}}a_1^\dag a_2^\dag$, where $a_i(a_i^\dag)$ being the annihilation (creation) operator and $g_i=(2\omega_i\nu_{\gamma_i})^{\frac{-1}{2}}$ with $\nu_{\gamma_i}$ the normalization volume of the photon mode for $i=1$ and 2 corresponding to the incident and scattered photon mode, respectively.  ${\bf k}_1$ (${\bf k}_2$) and $\boldsymbol{\epsilon}_1$ ($\boldsymbol{\epsilon}_2$) is the wave vector and polarization vector of the incident laser field (scattered photon mode), respectively.

\begin{figure*}[htbp]
\includegraphics{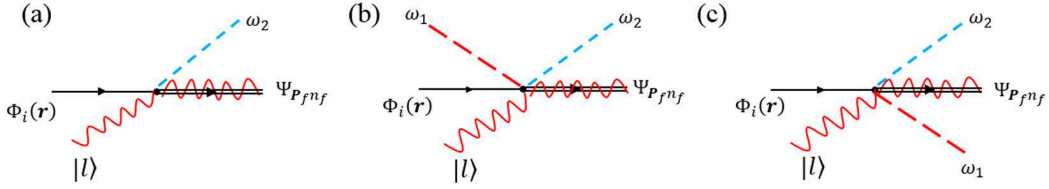}\centering
\caption{\label{fig1} Schematic for the one-step transition of Compton scattering by a bound electron. The single straight line represents the bound electron state, the wavy line represents the Fock state of the incident laser, the combination of wavy and double lines represents the Volkov state,
the red and blue dashed lines represent the scattered photons of frequency
$\omega_1$ and $\omega_2$ respectively, and the vertex denotes the transition operator $V_{22}$ in (\textbf{a}), $V_{21-}$ in (\textbf{b}), and $V_{21+}$ in (\textbf{c}).}
\end{figure*}

Figure~\ref{fig1} illustrates the corresponding schematic diagrams of the three terms in Eq.~(2), where we name $T_{AP}$ the laser-assisted electron-mode (LEM) transition shown in Fig.~\ref{fig1}(a), and name $T_{AA-}$ and $T_{AA+}$ the electron-assisted mode-mode (EMM) transitions shown in Fig.~\ref{fig1}(b) and (c) respectively. The LEM transition describes the process where the bound electron is ionized after absorbing several photons from the laser field, and at the same time, a photon of frequency $\omega_2$ is scattered, whereas the EMM transition describes a similar process except that a second photon of frequency $\omega_1$ is either absorbed ($T_{AA-}$) or emitted ($T_{AA+}$).

The matrix element of the LEM transition $T_{AP}$ can be written as
\begin{eqnarray}\label{eq13}
T_{AP}&=&\frac{e}{m}V_e^{-1/2}g_2\boldsymbol{\epsilon}_2^*\cdot[\textbf{P}_f+(u_p-q)\textbf{k}_1]\mathcal{J}_q(\zeta,\eta) \nonumber \\
&\times&\Phi_i(\textbf{P}_f+\textbf{k}_2+(u_p-q)\textbf{k}_1)\,,
\end{eqnarray}
where $\mathcal{J}_{q}(\zeta,\eta)=\sum_{m=-\infty}^\infty J_{-q-2m}(\zeta)J_m(\eta)$ is the generalized Bessel function, with $\zeta=2\sqrt{\frac{u_p}{m\omega_1}}\textbf{P}_f\cdot \boldsymbol{\epsilon}_1$ and $\eta=u_p/2$. $q=l-n_f$ denotes the number of photons transferred from the incident laser field during the NCS process. The matrix elements of the EMM transitions $T_{AA\pm}$ are given by
\begin{eqnarray}\label{eq14}
T_{AA-}&=&\frac{e^2}{m}V_e^{-1/2}\Lambda g_2\boldsymbol{\epsilon}_1\cdot \boldsymbol{\epsilon}_2^*\mathcal{J}_{q-1}(\zeta,\eta) \nonumber \\
&\times&\Phi_i(\textbf{P}_f+\textbf{k}_2+(u_p-q)\textbf{k}_1)
\end{eqnarray}
and
\begin{eqnarray}\label{eq15}
T_{AA+}&=&\frac{e^2}{m}V_e^{-1/2}\Lambda g_2\boldsymbol{\epsilon}_1^*\cdot \boldsymbol{\epsilon}_2^*\mathcal{J}_{q+1}(\zeta,\eta) \nonumber \\
&\times&\Phi_i(\textbf{P}_f+\textbf{k}_2+(u_p-q)\textbf{k}_1)\,,
\end{eqnarray}
where $\Lambda=\sqrt{\frac{u_p\omega_1m}{\alpha}}$ represents the half amplitude of the classical field in the limits of $g_1\rightarrow0$ and $l\rightarrow \infty$. In Eq. (3)-(5), the two-photon NCS processes correspond to $q=2$. Furthermore, the contribution of $T_{AA+}$ can be ignored, since it is much smaller than that of the other two terms under the present laser conditions.

The expression of the DDP for a Compton scattering process can be written as~\cite{karazija2013introduction}
\begin{equation}\label{eq16}
\frac{dW_{i\rightarrow f}}{d\omega_2d\Omega}=\int 2\pi|T_{fi}|^2\delta(E_i-E_f)\frac{\nu_{\gamma_2}}{(2\pi)^3}\frac{V_e}{(2\pi)^3}\omega_2^2d^3\!P_f\,,
\end{equation}
where $d\Omega$ is the differential solid angle of vector ${\bf k}_2$.
To analyze the results more clearly, we may rewrite the DDP by three parts:
\begin{equation}\label{eq20}
\frac{dW_{i\rightarrow f}}{d\omega_2d\Omega}=\frac{dW_{AP}}{d\omega_2d\Omega}+\frac{dW_{AA}}{d\omega_2d\Omega}+\frac{dW_{cro}}{d\omega_2d\Omega},
\end{equation}
where the three parts on the right-hand of the equation represent the contributions of EMM, LEM and their cross-term (CT). In the following, we will see how these terms affect the peak position of the DDP spectrum for two-photon NCS processes.

\section{The redshift}
\begin{figure*}[htpb]
\includegraphics{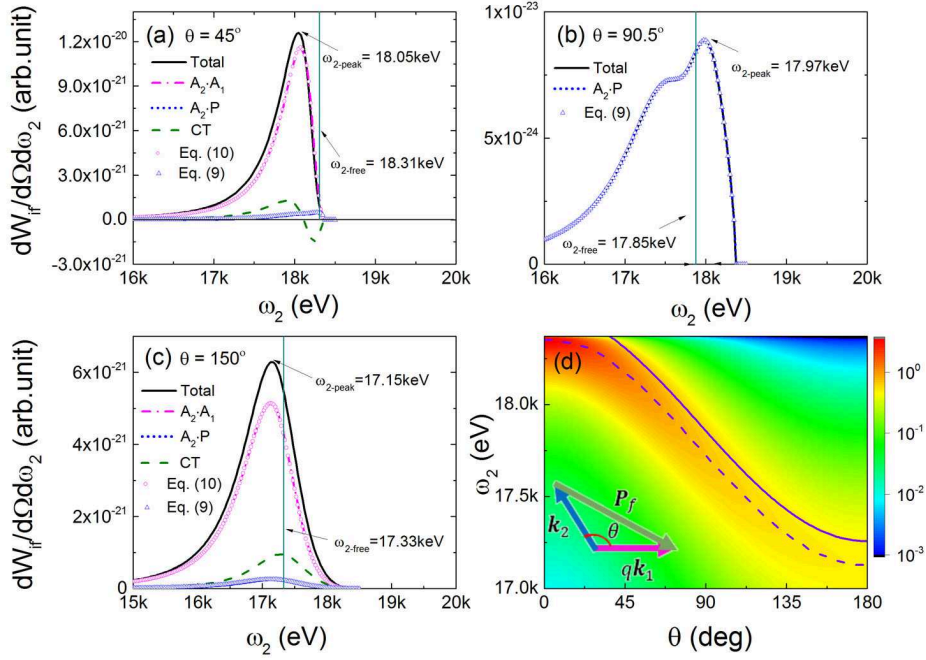}\centering
\caption{\label{fig2} The DDP as a function of the scattered photon energy $\omega_2$ at the scattering angle of $\theta=88^\circ$ (\textbf{a}), $\theta=90.5^\circ$ (\textbf{b}) and $\theta=150^\circ$ (\textbf{c}). The black solid lines represent the total DDP. The blue short-dotted lines, pink dash-dotted lines, and green dashed lines represent the DDP by LEM transition (Eq.~(\ref{eq13})), the DDP by EMM transition (Eq.~(\ref{eq14})) and the CT part of the total DDP, respectively. The blue triangles are for the DDP by Eq.~(\ref{eq18}) and the pink circles the DDP by Eq.~(\ref{eq19}). The vertical lines are $\omega_2$ predicted by Eq.~(\ref{eq17}). The scattered wave vector $\textbf{k}_2$ is fixed in the polarization plane defined by $\textbf{k}_1$ and $\boldsymbol{\epsilon}_1$. (\textbf{c}) The integral of $|\Phi_i(\textbf{P}_f+\textbf{k}_2+(u_p-q)\textbf{k}_1)|^2$ over $\textbf{P}_f$ as a function of $\omega_2$ and $\theta$. The dashed line denotes the scattered photon energy corresponding to the peaks at different scattering angles and the solid line represents the prediction by Eq.~(\ref{eq17}).}
\end{figure*}

 We now calculate the DDP for two-photon Compton scattering by a 1$s$ electron of Be atom, where the intensity of the laser field is ${\rm 4 \times 10^{20}~W/cm}^2$ and the photon energy is 9.25~keV. Figure~\ref{fig2} presents the DDP of the NCS at the scattering angles of $45^\circ$ (a), $90.5^\circ$ (b) and $150^\circ$ (c), where the wave vector of the scattered photon $\textbf{k}_2$ is fixed in the polarization plane of the incident laser field defined by $\textbf{k}_1$ ($z$-axis) and $\boldsymbol{\epsilon}_1$ ($x$-axis), i.e., the azimuthal angle $\phi=0^\circ$. In Fig.~\ref{fig2}(a)-(c), the vertical lines indicate the scattered photon energy predicted by the free-electron model~\cite{brown1964interaction,krebs2019time, venkatesh2020simulation}:
\begin{equation}\label{eq17}
\omega_2=\frac{q\omega_1}{1+\frac{q\omega_1}{m}(1-{\cos}\theta)}.
\end{equation}
It can be seen from Fig.~\ref{fig2} that the peak positions on the scattered spectra at $\theta=45^\circ$ and $150^\circ$ are red shifted with respect to the scattered photon energy predicted by the free-electron model, while this peak position at $\theta=90.5^\circ$ is blue shifted comparing with the scattered photon energy given by the free-electron model.

In order to explain the results shown in Fig. 2 (a)-(c), we present separately the contributions of LEM, EMM and CT by blue short-dotted lines (LEM), pink dash-dotted lines (EMM) and green dashed lines (CT) in Fig.~\ref{fig2}(a)-(c). One may find that the EMM dominates the contribution to DDP at the scattered angle $\theta=45^\circ$ and $150^\circ$ as shown in Fig.~2(a) and (c), and hence the redshift of the peak on the DDP spectra can be attributed to EMM transition. On the contrary, LEM transition dominates the contribution to the DDP at $\theta=90.5^\circ$, hence the buleshift of the peak on the DDP spectrum is due to LEM transition, as shown in Fig.~2(b). To find the reason of the shifts of these peaks, we may simplify Eqs.~(\ref{eq13})-(\ref{eq14}) by replacing the generalized Bessel functions $\mathcal{J}_{2}(\zeta,\eta)$ with $J_{-1}(\eta)+J_{-2}(\zeta)\approx C_1$ and $\mathcal{J}_{1}(\zeta,\eta)$ with $J_{-1}(\zeta)\approx C_2\textbf{P}_f\cdot \boldsymbol{\epsilon}_1$, because the values of $\zeta$ and $\eta$ are much smaller than 1 under the present laser conditions. Here the parameters $C_1$ and $C_2$ are constants determined by the laser conditions.  Therefore, the matrix element of LEM transition can be approximated as
\begin{eqnarray}\label{eq18}
T_{AP}\approx \frac{e}{m}V_e^{-1/2}g_2(P_f\cos{\theta}_{{\epsilon}_2}-q\textbf{k}_1\cdot\boldsymbol{\epsilon}_2^*) \nonumber \\
 \times C_1\Phi_i(\textbf{P}_f+\textbf{k}_2-q\textbf{k}_1)
\end{eqnarray}
with ${\theta}_{{\epsilon}_2}$ being the angle between  $\boldsymbol{\epsilon}_2^*$ and the electron momentum $\textbf{P}_f$. And the matrix element of EMM transition can be approximated as
\begin{eqnarray}\label{eq19}
T_{AA-}\approx \frac{e^2}{m}V_e^{-1/2}\Lambda g_2\boldsymbol{\epsilon}_1\cdot \boldsymbol{\epsilon}_2^*P_f\cos{\theta}_{{\epsilon}_1}C_2 \nonumber \\ \times\Phi_i(\textbf{P}_f+\textbf{k}_2-q\textbf{k}_1)
\end{eqnarray}
with ${\theta} _{{\epsilon}_1}$ being the angle between $\boldsymbol{\epsilon}_1$ and the electron momentum $\textbf{P}_f$. The DDP spectra by Eq.~(\ref{eq18}) and (\ref{eq19}) are shown by the triangles and circles respectively in Fig.~\ref{fig2}(a)-(c), where they agree with the corresponding numerical results.

We firstly consider the influence of the atomic wavefunction on the DDP spectra. By integrating the modular square of the wavefunction over $\textbf{P}_f$, we obtain the electron density distribution as a function of the scattered photon energy $\omega_2$ and the scattering angle $\theta$, as shown in Fig.~\ref{fig2}(d). It shows that the peaks presented by dashed line on the density distribution decreases with the scattering angle. This can be explained as follows: By analyzing the argument of the wavefunction, it can be found that the peak of the electron density distribution occurs at the momentum transfer $\textbf{P}_f=q\textbf{k}_1-\textbf{k}_2$, as illustrated by the the inset of Fig.~\ref{fig2}(d). Since this momentum transfer increases with the scattering angle, the energy gained by electron from the scattering process increase inevitably, leading to a decrease in the energy of the scattered photon. Moreover, one may find that the peaks of the density distribution are always redshifted relative to the free-electron model shown by the solid line, where the value of the redshift, in a range between 127~eV and 153~eV, is close to the binding energy of the 1s state of Be atom. This indicates that a bound electron may provide a redshift on the peak of the DDP spectrum by its binding energy.

Next, to explain the redshift of the peaks on the spectra in Fig.~2(a) and (c) further, we find from Eq.~(\ref{eq19}) that the DDP by EMM transition depends linearly on the value of the electron momentum $P_f$. In particular, the value of $P_f$ and thus the total DDP decrease as the scattered photon energy increases due to the energy conservation. As a result, an obvious redshift of the DDP spectrum is formed, which is mainly caused by the dependence of DDP on the ionized electron momentum and bound-state wavefunction, as shown in Fig.~2(a) and (c). At last, to understand the blueshift of the peaks on the DDP spectrum at $\theta=90.5^\circ$ shown in Fig.~2(b), we analyse Eq.~(\ref{eq18}) and find that the term $P_f\cos{\theta}_{{\epsilon}_2}-q\textbf{k}_1\cdot\boldsymbol{\epsilon}_2^*$ in Eq.~(\ref{eq18}) plays a crucial role on the shift of the peaks, where the value of this term increases at around $\theta=90^\circ$ with the scattered photon energy, leading to the blueshift of peak on the DDP spectrum.

\begin{figure*}[htpb]
\includegraphics{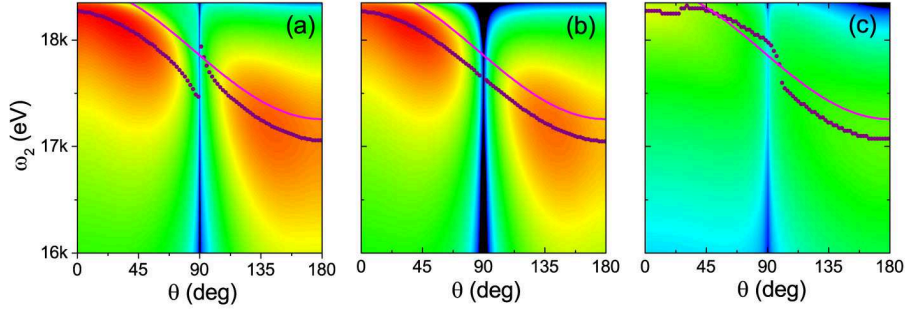}\centering
\caption{\label{fig3} Angle-resolved energy spectra of two-photon NCS by Be. The DDP is show as a function of the scattering angle $\theta$ and the scattered photon energy $\omega_2$ at $\phi=0^\circ$. The total DDP, the DDP due to EMM, and the DDP due to LEM are shown in (\textbf{a}), (\textbf{b}) and (\textbf{c}), respectively. The solid lines represent $\omega_2$ determined by Eq.~(\ref{eq17}). The dots denote the peaks at different scattering angles for the total DDP, the DDP of EMM term and the DDP of LEM term, respectively.}
\end{figure*}

We then consider how the EMM and LEM transitions influence the total DDP for different scattering angles. Fig.~3(a) presents the total DDP spectrum as a function of scattered photon energy $\omega_2$ and scattering angle $\theta$ at $\phi=0^\circ$, where the dots show the peaks of the DDP spectra and the solid lines predict the scattered photon energy by the free-electron model. One may find that, except for a narrow region of the scattering angle around $90^\circ$, the peak value of the DDP spectra decreases with the scattering angle and it is always red shifted relative to the scattered photon energy of the free-electron model with a maximum redshift of 400~eV. Especially, the peak at $\theta=90^\circ$ is blue shifted comparing with the scattered photon energy by the free-electron model. To explain these results, we present the DDP spectrum by EMM and LEM transitions in Fig.~3(b) and (c), respectively. The dots in Fig.~3(b) and (c) show the peaks of the DDP spectra and the solid lines show the prediction of the scattered photon energy by the free-electron model. On the one hand, we may find that the the peaks of the DDP spectra by EMM decrease with the scattering angle and is always red shifted comparing with the value of the free-electron model, as shown in Fig.~3(b). In additionally, the DDP spectrum of EMM transition presents a dip at about $\theta=90^\circ$, which is also confirmed in Fig.~2(b). This is because that the DDP of EMM transition is proportional to $|\boldsymbol{\epsilon}_1\cdot \boldsymbol{\epsilon}_2^*|^2$, which is zero at $\theta=90^\circ$ due to the scattering geometry of $\boldsymbol{\epsilon}_2\perp \boldsymbol{\epsilon}_1$ since $\textbf{k}_2 \perp  \textbf{k}_1$ and $\textbf{k}_2\parallel \boldsymbol{\epsilon}_1$. On the other hand, it can be found that the peaks of the DDP spectra by LEM presents a complex situation, where the peaks around $\theta=90^\circ$ are blue shifted comparing with the prediction of the free-electron model and are red shifted at other scattering angles. Furthermore, comparing Fig.~3(a) with (b) and (c), we may find that the EMM transition dominates the contributions on the DDP spectrum, except for the region around $\theta=90^\circ$ where the LEM dominates the total DDP spectrum. Therefore, the red shift of the peaks on the total DDP spectrum is attributed to the EMM transition, i.e., the dependence of DDP on the ionized electron momentum and bound-state wavefunction. Additionally, although the forward scattering is not given in the work of Fuchs \emph{et al.}~\cite{fuchs2015anomalous}, our results show that the redshift phenomenon still exists significantly in the forward scattering.

\begin{figure*}[htpb]
\centering\includegraphics[width=13cm]{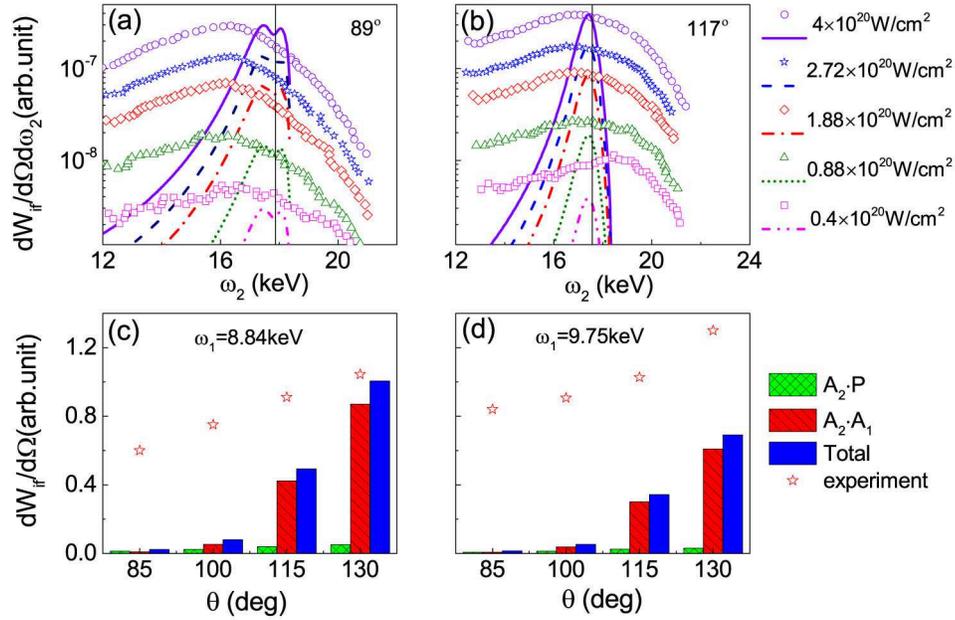}
\caption{\label{fig4} Comparison between the DDP of theory and experiment. The DDP for two-photon NCS by Be at $\theta=89^\circ$ (\textbf{a}) and $\theta=117^\circ$ (\textbf{b}). The curves and geometric figures represent, respectively, the theoretical and experimental values under different laser intensities. The incident photon energy $\omega_1$ is 9.25 keV. (\textbf{c})-(\textbf{d}) Comparison of SDP between theory and experiment at different scattering angles at $\omega_1=8.84$ keV (c) and  $\omega_1=9.75$ keV (d). The stars represent the experimental data points. Note: the theoretical values in the same graph are magnified by the same multiple.}
\end{figure*}

We now qualitatively compare our NCS spectra with the experimental results \cite{fuchs2015anomalous} by using a magnification factor $\delta$, where the DDP of two-photon NCS is shown in Fig.~\ref{fig4} at $\theta=89^\circ$ with $\delta=2\times10^{16}$ (a) and $\theta=117^\circ$ with $\delta=1.11\times10^{14}$ (b). In Fig.~\ref{fig4}(a)-(b), the curves and geometric symbols represent, respectively, the theoretical and experimental results under various laser intensities. Compared with the scattered photon energy predicted by free-electron model at $\theta=89^\circ$ and $\theta=117^\circ$, the peak energy obtained from our theory red shifts 393 eV and 155 eV, respectively. Besides, it shows that the change of theoretical DDP with laser intensity agrees with the experimental results, which originates from the second-order nonlinear effects of laser intensity.

By integrating the DDP over the scattered photon energy, we obtain the corresponding single differential probability (SDP) shown in Fig.~\ref{fig4}(c) and (d) with $\delta=1.4\times10^{17}$ for the case of $\omega_1=8.84$~keV (c) and  $\omega_1=9.75$~keV (d). It shows that the SDP increases as the scattering angle increases, which is in agreement with the experimental results displayed by the red stars in graphs. Moreover, the SDP from EMM and LEM transitions are also presented in Fig. 4(c) and (d). One may find that the SDP contributed by LEM transition is insensitive to the scattering angle, whereas the SDP by EMM transition increases with the scattering angle. Therefore, the dependence of the total SDP on the scattering angle can be mainly attributed to the EMM transition.


\section{Conclusion}
We have extended the frequency-domain theory to investigate the NCS of two X-ray photons by an atom. Our theoretical results are in qualitative agreement with the experimental results of Ref.~\cite{fuchs2015anomalous} and thus the underlying physical mechanism for the nonlinear scattering process, i.e., the observed anomalous redshifts are attributed to the atomic binding potential and the momentum transfer from the incident photons to the electron during the collision. Our results have demonstrated that the redshift can be observed in in both forward and backward directions. All these findings promote significantly the understanding of the nonlinear scattering processes of bound electrons in X-ray laser fields.
\begin{backmatter}
\bmsection{Funding}
National Natural Science Foundation of China under Grant Nos. 12074418, 11774411 and 11834015. ZCY was supported by the NSERC of Canada.

\bmsection{Acknowledgments}
We thank all the members of SFAMP club for helpful discussions. S.S thanks D. Krebs and M. Fuchs for helpful discussions.

\bmsection{Disclosures}
The authors declare no conflicts of interest.

\bmsection{Data availability} Data underlying the results presented in this paper are not publicly available at this time but may be obtained from the authors upon reasonable request.

\end{backmatter}

\bibliography{cankao}  






\end{document}